\documentclass[twocolumn,superscriptaddress,nofootinbib]{revtex4-1}

\usepackage[utf8]{inputenc} 
\usepackage[T1]{fontenc}    
\usepackage{hyperref}       
\usepackage{url}            
\usepackage{booktabs}       
\usepackage{amsfonts}       
\usepackage{nicefrac}       
\usepackage{microtype}      
\usepackage{lipsum}
\usepackage{graphicx}
\usepackage{amsmath}
\usepackage{float}

\usepackage[table]{xcolor}
\usepackage{color}
\usepackage{colortbl}
\usepackage{multirow}

\definecolor{midgreen}{rgb}{0.52, 0.73, 0.4}

\graphicspath{ {figures/} }

\begin{document}

\title{Smooth matching of $\hat{q}$ from hadronic to quark and gluon degrees of freedom}

\author{Emma McLaughlin}
\affiliation{Department of Physics, Columbia University, 538 West 120th Street, New York, NY 10027, USA}
\author{Jacob Rose}
\affiliation{Department of Physics, Dr. Karl-Remeis Observatory  - Astronomical Institute,
Bamberg D-96049, Germany}
\author{Paolo Parotto}
\affiliation{University of Wuppertal, Department of Physics, Wuppertal D-42097, Germany}
\author{Claudia Ratti}
\affiliation{Department of Physics, University of Houston, Houston, TX, USA  77204}
\author{Jacquelyn Noronha-Hostler}
\affiliation{Illinois Center for Advanced Studies of the Universe, Department of Physics, University of Illinois at Urbana-Champaign, Urbana, IL 61801, USA}

\begin{abstract}
One of the key signatures of the Quark Gluon Plasma (QGP) is the energy loss of high momentum particles as they traverse the strongly interacting medium. The energy loss of these particles is governed by the jet transport coefficient $\hat{q}/T^3$, wherein it has been thought that there is a large jump 
as the system transitions between the hadron gas and Quark Gluon Plasma phases.  Here we calculate $\hat{q}/T^3$ within the Hadron Resonance Gas (HRG) model with the particle list PDG16+ and find that, if one incorporates the experimental error in the hadronic calculation of $\hat{q}/T^3$ and assumes a higher pseudo-critical temperature, then a smooth transition from the hadron gas phase into the Quark Gluon Plasma phase is possible. We also find a significant enhancement in $\hat{q}/T^3$ at finite baryon chemical potential and find issues with the relationship between the shear viscosity and the jet transport coefficient within a hadron gas phase. 
\end{abstract}
 

\maketitle

\section{Introduction}
While jets are primarily thought of as a tool that probe the early stages of heavy-ion collisions \cite{Adil:2005qn,Andres:2019eus,Connors:2017ptx}, the jet transport coefficient $\hat{q}$ is influenced by the contributions of $\hat{q}_{HRG}$ from the hadron resonance gas as well. In the well-know JET Collaboration paper \cite{Burke:2013yra} a summary of $\hat{q}(T)$ calculations is shown, using a variety of models bench-marked against the same medium.  A primary conclusion from that work was that there is a large jump in $\hat{q}$ from the hadron gas phase at low temperature where $\hat{q}/T^3\lesssim 1$, up to the QGP phase where $\hat{q}/T^3\sim 4.5$.  The argument for this behavior is that the QGP phase is strongly interacting, wherein high $p_T$ partons would lose a significant amount of energy (hence a large $\hat{q}/T^3$), whereas the hadron resonance gas phase is weakly interacting and the high-$p_T$ partons would pass through it
largely unaffected. 

However, our current picture of the QGP/hadron resonance gas phase transition has evolved significantly over the years and this may no longer hold.  For instance, we know that the QGP/hadron resonance gas transition is a smooth cross-over with no critical point/first-order phase transition anywhere near the $\mu_B=0$ range that is probed at the LHC and top RHIC energies \cite{Aoki:2006br,Borsanyi:2010cj,Borsanyi:2010bp,Bazavov:2014pvz,Bazavov:2017tot,Bazavov:2017dus,Mukherjee:2019eou,Borsanyi:2020fev}. Additionally, this implies that there may well be different hadronization and freeze-out temperatures for light vs. strange hadrons 
\cite{Bellwied:2013cta,Bellwied:2019pxh,Noronha-Hostler:2016rpd,Bellwied:2018tkc,Bluhm:2018aei,Alba:2020jir,Flor:2020fdw}. Furthermore, there is no reason to believe that the inflection temperatures of different transport coefficients should occur at a single temperature either.  In fact, in a bottom-up picture from holography \cite{Rougemont:2017tlu} that has been tuned to precisely fit lattice QCD \cite{Critelli:2017oub,1848212} (and even make predictions for high-order susceptibilities), it was found that there is a large range ($T_{pc}\sim 130-200$ MeV) of pseudo-critical temperatures for 
the relevant transport coefficients. In this work, it was found that the inflection point for $\hat{q}/T^3$ is located around $T_{pc}\sim 195$ MeV, which implies that calculations of $\hat{q}/T^3$ in a hadron resonance gas model may be relevant up to high temperatures.  Additionally, it also implies that it may be possible to smoothly transition from $\hat{q}/T^3$ in a hadron resonance gas picture to a QGP picture.  

In this manuscript, we calculate $\hat{q}/T^3$ within a hadron resonance gas 
model and assume a higher $T_{pc}$ as well as incorporate all experimental error bars from the nuclear saturation density and the jet transport parameter for a quark at the center of a nucleus.  Once these two effects are incorporated in our calculations, we find that $\hat{q}/T^3$ can smoothly connect to the extracted $\hat{q}/T^3$ from the JET collaboration \cite{Burke:2013yra}. We then extend this work to finite baryon densities where $\hat{q}/T^3$ is approximately double in value at $\mu_B\sim400$ MeV. We  also find that large differences exist at $\mu_B>400$ MeV, when excluded volume effects are included in the calculation. 
Finally, we test the quasi-empirical relationship between the shear viscosity and the jet transport coefficient derived in Ref. \cite{Majumder:2007zh} and find that it does not seem to hold in the hadronic phase.

\section{Hadron Resonance Gas Model}

In the hadron resonance gas model, one can calculate the pressure and energy density of a system of hadrons, assuming that they are point-like particles, according to the following formulas:
\begin{widetext}
\begin{eqnarray}
    \frac{p(T,\mu_B,\mu_S,\mu_Q)}{T^{4}} &=&  \sum\limits_{j}(-1)^{B_{j}+1} \frac{g_{j}}{2\pi^{2}}\int\limits_{0}^{\infty} p^{2}\ln{[1+(-1)^{B_{j}+1}e^{(-\sqrt{p^{2}+mj^{2}}+ B_{j}\mu_{B}+S_{j}\mu_{S}+Q_{j}\mu_{Q})/T}]} dp\label{eqn:hrgp}\\
   \frac{ e(T,\mu_B,\mu_S,\mu_Q)}{T^4}&=&  \sum\limits_{j}\frac{g_{j}}{2\pi^{2}}\int\limits_{0}^{\infty} \frac{p^{2}\sqrt{p^{2}+m_j^{2}}}{e^{(-\sqrt{p^{2}+m_j^{2}}+ B_{j}\mu_{B}+S_{j}\mu_{S}+Q_{j}\mu_{Q})/T}+(-1)^{B_{j}+1}} dp\label{eqn:hrge}
\end{eqnarray}
\end{widetext}
where $g_j$ is the degeneracy of each hadron, $m_j$ is the mass, and $B_j$, $S_j$, and $Q_j$ are the baryon number, strangeness and electric charge carried by each hadron.  Additionally, $\mu_B$, $\mu_S$, and $\mu_Q$ are the corresponding chemical potentials for each conserved charge. The sums run over all possible baryons and mesons belonging to a given particle list. We will discuss our particle list choices below.

In order to calculate $\hat{q}_{HRG}$ we use Eq.\ (9) from Ref.~\cite{Burke:2013yra} 
\begin{equation}\label{eqn:qhat}
\hat{q}_{HRG}(T)=\frac{\hat{q}_N}{\rho_N}\left[\frac{2}{3}\rho_M(T)+\rho_B(T)\right]
\end{equation}
where $\rho_N\sim (0.15-0.17) fm^{-3}$ is the nucleon density in the center of a large nucleus and  the jet transport parameter for a quark at the center of a large nucleus, $\hat{q}_N\sim (0.024 \pm 0.008)$ GeV$^2$/fm, is extracted \cite{Wang:2009qb} from HERMES data \cite{Airapetian:2007vu}.  If we take the range of error-bars into account, the prefactor in the above equation lies in the range $\hat{q}_N/\rho_N=2.4-5.8$. Throughout the rest of this section, the bands shown in the plots for $\hat{q}$ take this variation into account. The factor of $2/3$ in the above equation assumes that mesons are composed of 2 quarks and baryons contain 3 quarks. While this assumption is certainly true for valence quarks (namely in the high $x$ regime), at the $x$ and $Q^2$ values probed in heavy-ion collisions at top RHIC and LHC energies sea quarks play a significant role (see e.g. Fig. 19 from  HERAPDF1.0 \cite{Aaron:2009aa}). Thus, it is not clear whether this factor of 2/3 should still remain, but we leave it in the calculation for lack of a better definition.

In Eq.~(\ref{eqn:qhat}), $\rho_M(T)$ is the total particle density of mesons and $\rho_B(T)$ is the total particle density of baryons, both calculated within the HRG model.  The total particle densities can be calculated using the partial pressures of mesons $p_M$ and baryons $p_B$, such that $\rho_M=p_M/T$ and $\rho_B=p_B/T$.
The partial densities are then defined as:
\begin{widetext}
\begin{eqnarray*}
  \rho_M(T,\mu_B,\mu_S,\mu_Q) &=& \sum\limits_{j\in M} \frac{-g_{j}T^3}{2\pi^{2}}\int\limits_{0}^{\infty}p^2 \left[ \exp \left(\frac{\sqrt{p^2+m_i^2}}{T}-\frac{(B_{j}\mu_{B}+S_{j}\mu_{S}+Q_{j}\mu_{Q}}{T}\right)+1\right]^{-1} dp\label{eqn:hrgpm}\\
   \rho_B(T,\mu_B,\mu_S,\mu_Q) &=& \sum\limits_{j\in B} \frac{g_{j}T^3}{2\pi^{2}}\int\limits_{0}^{\infty}p^2 \left[ \exp \left(\frac{\sqrt{p^2+m_i^2}}{T}-\frac{(B_{j}\mu_{B}+S_{j}\mu_{S}+Q_{j}\mu_{Q}}{T}\right)-1\right]^{-1} dp
   \label{eqn:hrgpb}
\end{eqnarray*}
\end{widetext} 
Both quantities will be influenced by the number of resonances in our system, especially close to the cross-over phase transition. In this paper we compare two different lists from the Particle Data Group, one from 2005 (PDG05) and another developed in Ref.~\cite{Alba:2017mqu} from 2016 that includes all *-**** states (PDG16+).


We will also compare the ideal HRG $\hat{q}/T^3$ to that from an excluded volume picture.  Since the prefactor $\frac{\hat{q}_N}{\rho_N}$ is unchanged in an excluded volume picture, we only need to include excluded volume effects in the total particle density.  First we define the ideal total density (where mesons are weighted by $\frac{2}{3}$):
\begin{equation}
    \rho_{q}\equiv\frac{2}{3}\rho_{M}+\rho_{B}.
\end{equation}
Then the excluded volume version follows from the transcendental equation
\begin{equation}
 \rho_{v,q}(T,\mu_i)= \rho_q(T,\mu_i) \exp(-v\rho_{v,q}(T,\mu_i)),
\label{vdwpressure}
\end{equation}
which can be easily solved using a Lambert W function
\begin{equation}
 \rho_{v,q}(T,\mu_i)= \frac{1}{v}W(v \rho_q(T,\mu_i))\,.
\label{vdwpressuresolution2}
\end{equation}
Finally, the excluded volume $\rho_{v,q}(T,\mu_i)$ can be used to calculate the excluded volume
version of $\hat{q}$
\begin{equation}
    \hat{q}_{v}(T,\mu_i)=\frac{\hat{q}_N}{\rho_N}\rho_{v,q}(T,\mu_i).
\end{equation}

\section{Results: Continuous $\hat{q}/T^3 (T)$}

\begin{figure}
\begin{center}
\includegraphics[width=0.5\textwidth]{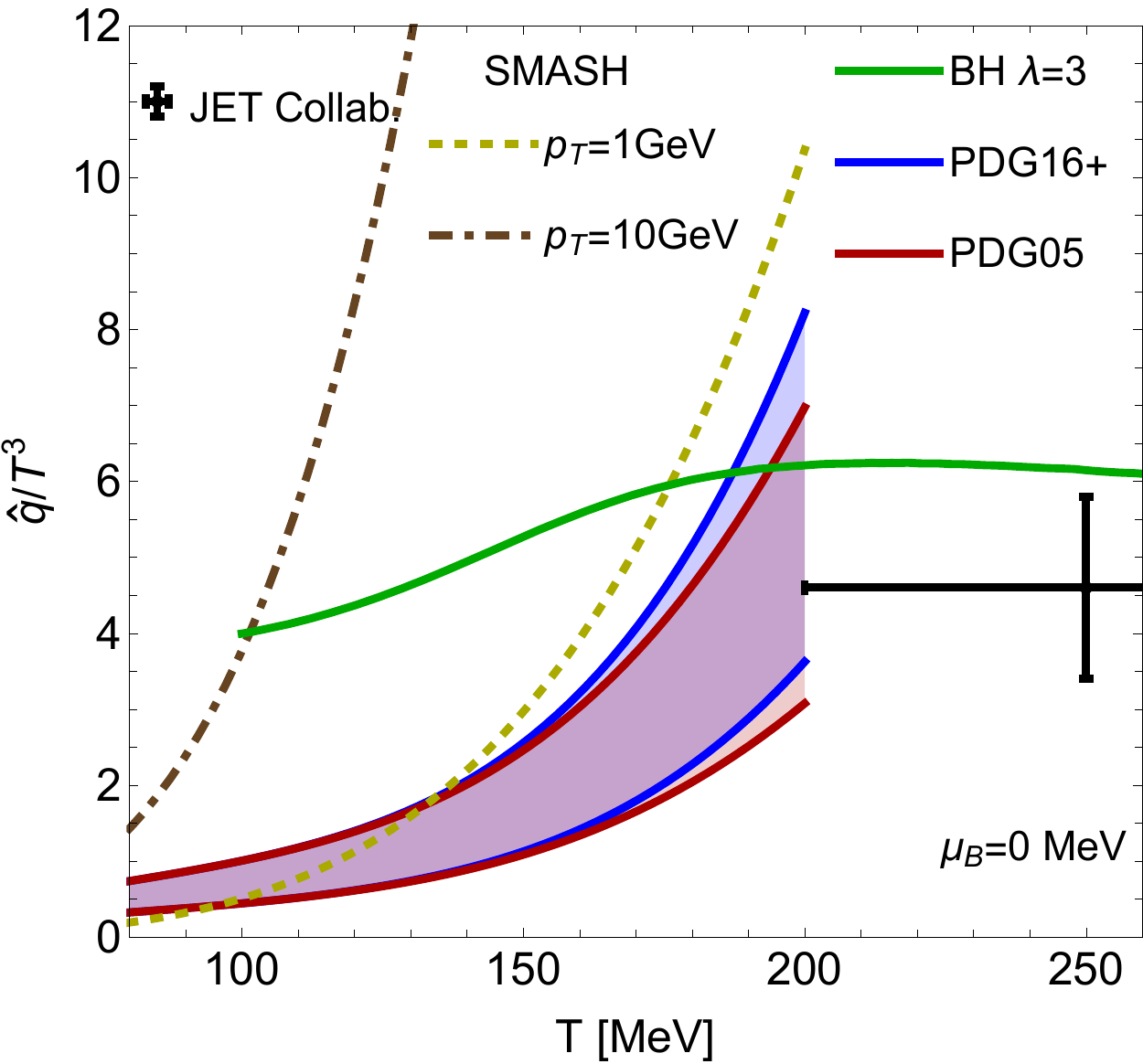} 
\end{center}
\caption{
$\hat{q}_{HRG}(T)$ calculations within the HRG model for PDG05 (red band) and PDG16+ (blue band). The band corresponds to the uncertainty in $\hat{q}_N/\rho_N$. The holography curve is taken from Ref.~\cite{Rougemont:2017tlu} assuming $\lambda=g^2*N_c=3$.  The JET collaboration point is from Ref.~\cite{Burke:2013yra}. 
} 
\label{fig:qhat}
\end{figure}
In Fig.\ \ref{fig:qhat} we show a comparison between $\hat{q}_{HRG}(T)$ for the PDG05 and PDG16+, calculated in the ideal HRG model. We find, indeed, that there is a small increase in $\hat{q}_{HRG}(T)$ close to the phase transition when more resonances are included in the hadronic spectrum. However, the inclusion of additional resonances has only a small effect compared to the uncertainties that remain in $\hat{q}_N/\rho_N$.  In fact, 
once one considers the entire error band from $\hat{q}_N/\rho_N$, a continuous transition from $\hat{q}_{HRG}(T)$ into $\hat{q}_{QGP}(T)$ is possible. 
Comparing our results to Fig. 10 in Ref.~\cite{Burke:2013yra}, we find a much smoother transition of $\hat{q}_{HRG}(T)$ into the QGP phase, which may affect energy loss models. Our results match quite well the JET collaboration's extracted RHIC value of $\hat{q}/T^3=4.6\pm 1.2$ if we assume the peak of $\hat{q}$ occurs at $T_{pc}\sim 200$ MeV, as predicted from holography \cite{Rougemont:2017tlu}.  As far as we are aware, the temperature dependence of $\hat{q}_{HRG}(T)$ at the cross-over has not yet been investigated in energy loss models but it would be interesting to explore a smoother transition vs. the more abrupt one that was originally seen in Ref.~\cite{Burke:2013yra}. 

The consequences of a smoother $\hat{q}(T)$ would imply that it is possible to still lose energy even when hadronic degrees of freedom begin to appear.  Thus, one should systematically study a variety of decoupling temperatures from the medium, as has been done in Refs.~\cite{Betz:2016ayq,Nahrgang:2016lst,Prado:2016szr,Andres:2019eus,Katz:2019fkc}.

We note that there are subtleties when determining $\hat{q}$ in a hadronic transport model noted in Ref.~\cite{Dorau:2019ozd}, that go beyond the purpose of this work.  In such an approach, one can better study the influence of $2\leftrightarrow 2$ interactions and the influence of heavier resonances may play a larger role, but we leave that study for a future work. One must first incorporate the additional resonances from the PDG16+ into SMASH, which is beyond the scope of this paper.

\begin{figure}
\begin{center}
\includegraphics[width=0.5\textwidth]{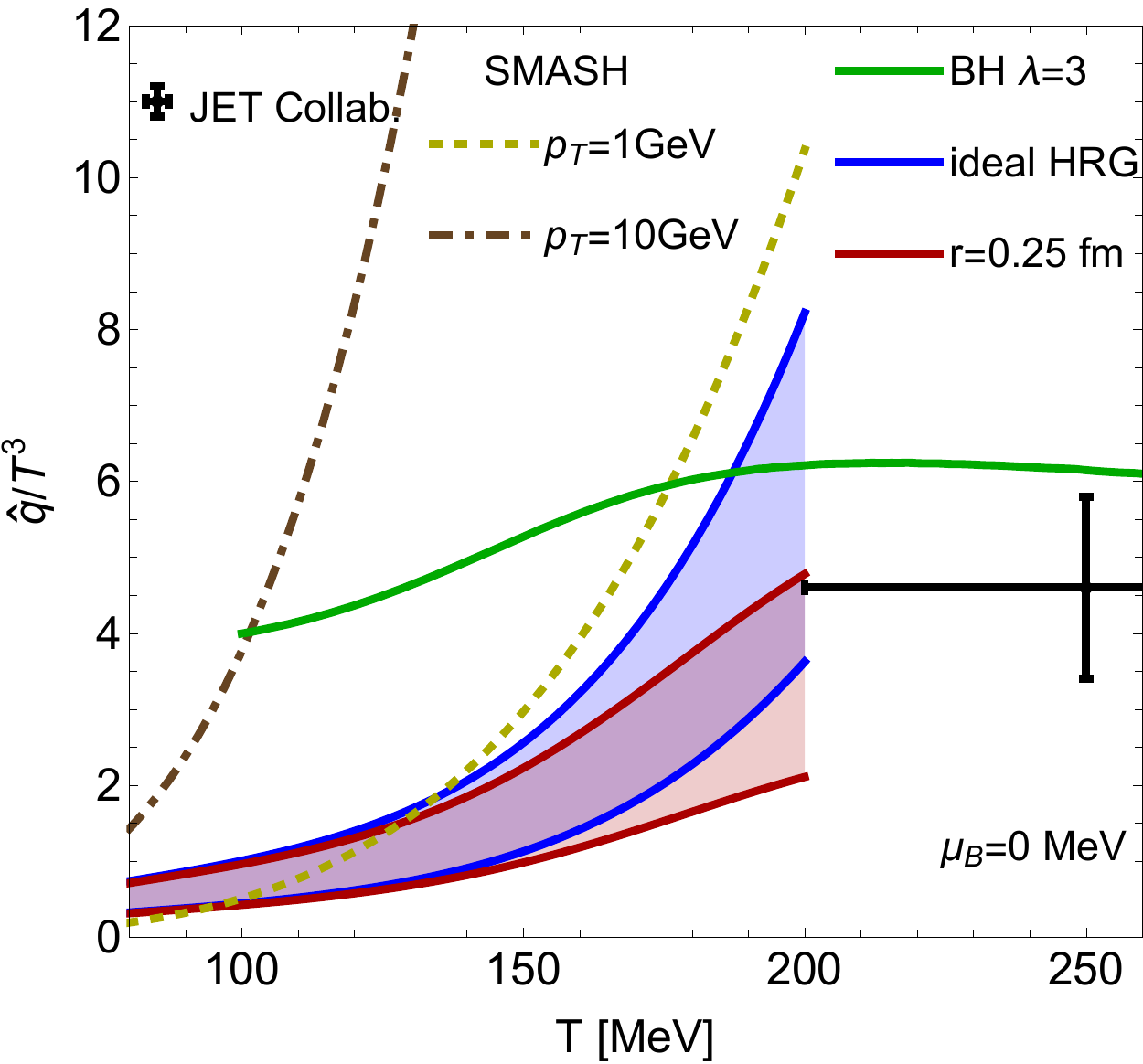} 
\end{center}
\caption{
$\hat{q}_{HRG}(T)$ calculations for PDG16+ with an ideal HRG vs. excluded volume with $r=0.25$ fm, where the band takes into account the error in $\hat{q}_N/\rho_N$. The holography curve is taken from Ref.~\cite{Rougemont:2017tlu} assuming $\lambda=g^2*N_c=3$.  The JET collaboration point is from Ref.~\cite{Burke:2013yra} and the SMASH calculations from Ref.~\cite{Dorau:2019ozd}. 
} 
\label{fig:qhatEX}
\end{figure}

We now proceed to study the effect of an excluded volume approach in the hadron resonance gas model.  In Ref.~\cite{McLaughlin:2021dph} it was shown that the maximum radius for a volume with the PDG16+ is $r=0.25$ fm, which corresponds to the maximal effect on $\hat{q}$. We find that this correction leads to a suppression of $\hat{q}/T^3$ at high temperatures, but we still find a reasonable match to the result of the JET collaboration, as shown in Fig.\ \ref{fig:qhatEX}. While in the case of the ideal HRG model the difference in $\hat{q}(T)/T^3 $ between PDG16+ and PDG05 was not very significant, we point out that the PDG05 barely supports any excluded volume at all when comparing to lattice QCD results (as shown in Ref.~\cite{McLaughlin:2021dph}), so that only the PDG16+ list would allow any effects from excluded volume. Thus, for the rest of the paper we only show PDG16+ results, for which we consider the ideal and excluded volume approaches.

In Fig.\ \ref{fig:qhatEX}, we compare ideal and excluded volume results for $\hat{q}(T)/T^3$ to both the holographic result from Ref.~\cite{Critelli:2017oub} and the formulas from hadronic transport depending on $p_T$ \cite{Dorau:2019ozd}.  In the case of holography, the results are given in terms of $\hat{q}(T)/T^3 /\sqrt{\lambda}$, where $\lambda= 4\pi\alpha N_c = N_c g^2$ is the t`Hooft coupling.  For illustrative purposes we assume $\lambda=3$, which seems a reasonable estimate.  Additionally, we also compare to SMASH for $p_T=1$ GeV and $p_T=10$ GeV (with the caveat that in Ref.~\cite{Dorau:2019ozd} it is pointed out that their calculations might not be directly equivalent to $\hat{q}$) and find that $p_T=1$ GeV is much closer to our estimate.  We note that, in the HRG model formula from this paper, there is no explicit $p_T$ dependence of a jet, so it is difficult to make an apples-to-apples comparison.

\section{Results: Energy Loss at the beam energy scan - $\hat{q}(T,\mu_B)$}

At lower beam energies, generally one does not expect signatures of energy loss because of the shorter lifetime of the QGP (e.g. see Fig. 9 in Ref.~\cite{Auvinen:2013sba} for lifetime estimates). However, recent measurements from STAR may indicate jet energy loss \cite{Adamczyk:2017nof}. To understand energy loss better at low beam energies, we can easily extend Eq.\ (\ref{eqn:qhat}) to finite $\left\{\mu_B,\mu_S,\mu_Q\right\}$ within a hadron resonance gas picture.  

\begin{figure}
\begin{center} 
\includegraphics[width=0.5\textwidth]{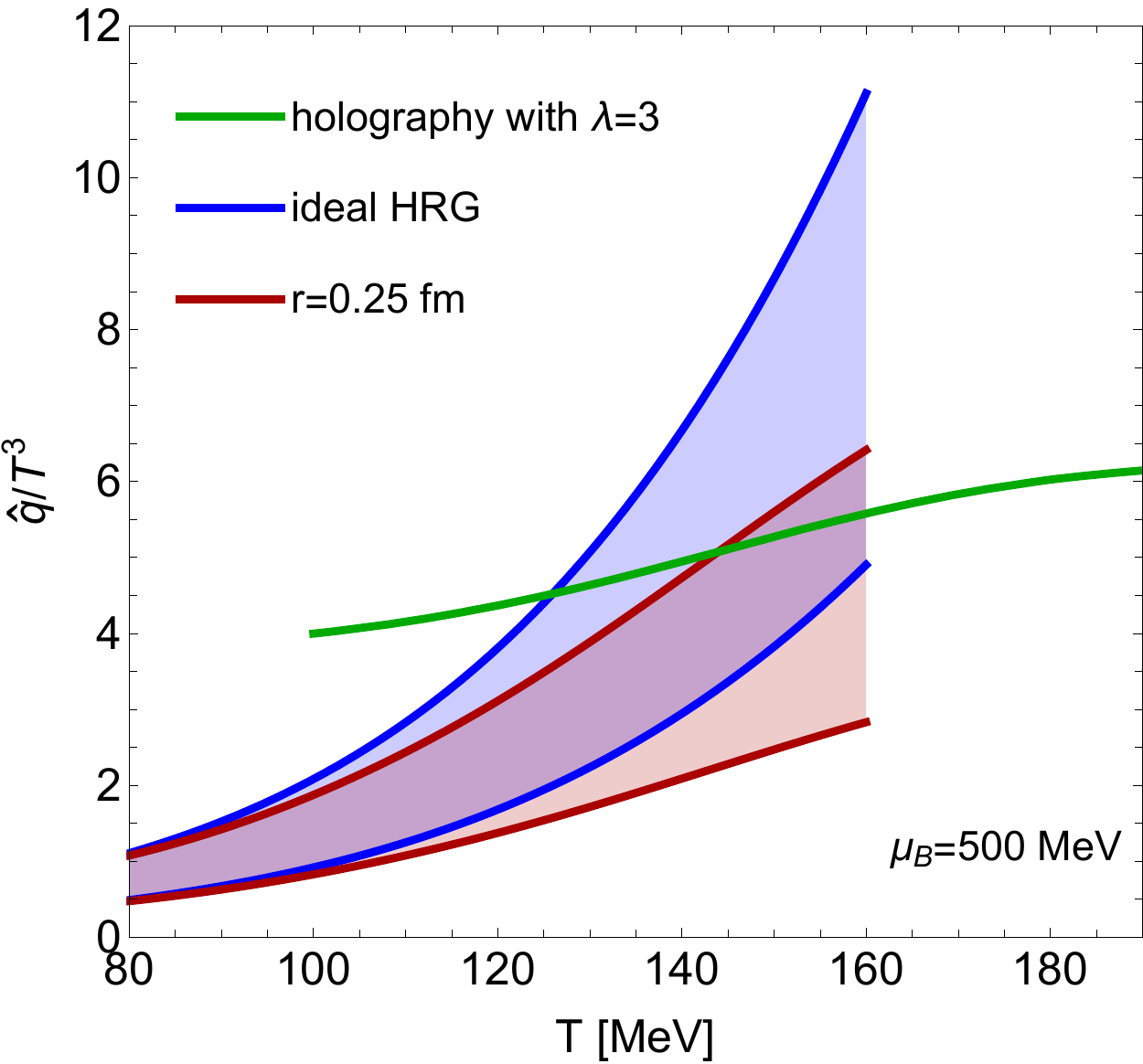} 
\end{center}
\caption{
$\hat{q}_{HRG}(T)$ at $\mu_B=500$ MeV calculations within the hadron resonance gas for  PDG16+ with an ideal HRG vs. excluded volume approach where the band corresponds to the error in $\hat{q}_N/\rho_N$. The holography curve is taken from Ref.~\cite{Rougemont:2017tlu} assuming $\lambda=g^2*N_c=3$.
} 
\label{fig:qhat500}
\end{figure}

In Fig.\ \ref{fig:qhat500}, we show $\hat{q}_{HRG}(T,\mu_B=500~\mathrm{MeV})/T^3$ and find a considerable enhancement, compared to the result at zero chemical potential discussed before.  Additionally, since it is expected that the transition temperature decreases at finite $\mu_B$, we only plot $\hat{q}_{HRG}/T^3$ up to $T_{pc}=160$ MeV, using holography at large $\mu_B$ as guidance \cite{Rougemont:2017tlu}. Since  $\hat{q}_{HRG}/T^3$ is nearly double than the one at $\mu_B=0$, one can conclude that even with hadronic degrees of freedom, one can anticipate energy loss at low beam energies, which is consistent with results from STAR \cite{Adamczyk:2017nof}. This is, of course, not a comment on the production of jets, which we would intuitively expect to be suppressed at low beam energies, but we are unaware of any calculations that show this explicitly.

In order to have a better connection to experiments, we use isentropes to demonstrate a possible trajectory across the QCD phase diagram. Here we calculate $\hat{q}/T^3$ along the isentropes extracted from the BSQ EoS from Ref.~\cite{Noronha-Hostler:2019ayj}, because it also uses the PDG16+ (note these isentropes were only extracted for the ideal HRG so we do not show the excluded volume results in the following). We show a comparison between the simplistic QCD EoS that is only 2D i.e. $\left\{T,\mu_B\right\}$ vs. the realistic one that contains all 3 conserved charges within heavy-ion collisions: baryon number $B$, strangeness $S$, and electric charge $Q$.  On average one expects strangeness neutrality i.e. $\langle \rho_S\rangle =0$ and the fraction of protons to neutrons to be fixed by the type of nuclei collided such that $\langle \rho_Q\rangle\sim 0.4 \langle \rho_B\rangle$. We note that the picture is, of course, much more complex once one considers local charge fluctuations and out-of-equilibrium effects \cite{Shen:2017ruz,Feng:2018anl,Fotakis:2019nbq,Martinez:2019rlp,Martinez:2019jbu,Dore:2020jye}, but this at least provides us with a general idea of the consequences of just considering baryon number conservation vs. the full BSQ EoS.

\begin{figure}
    \centering
    \includegraphics[width=\linewidth]{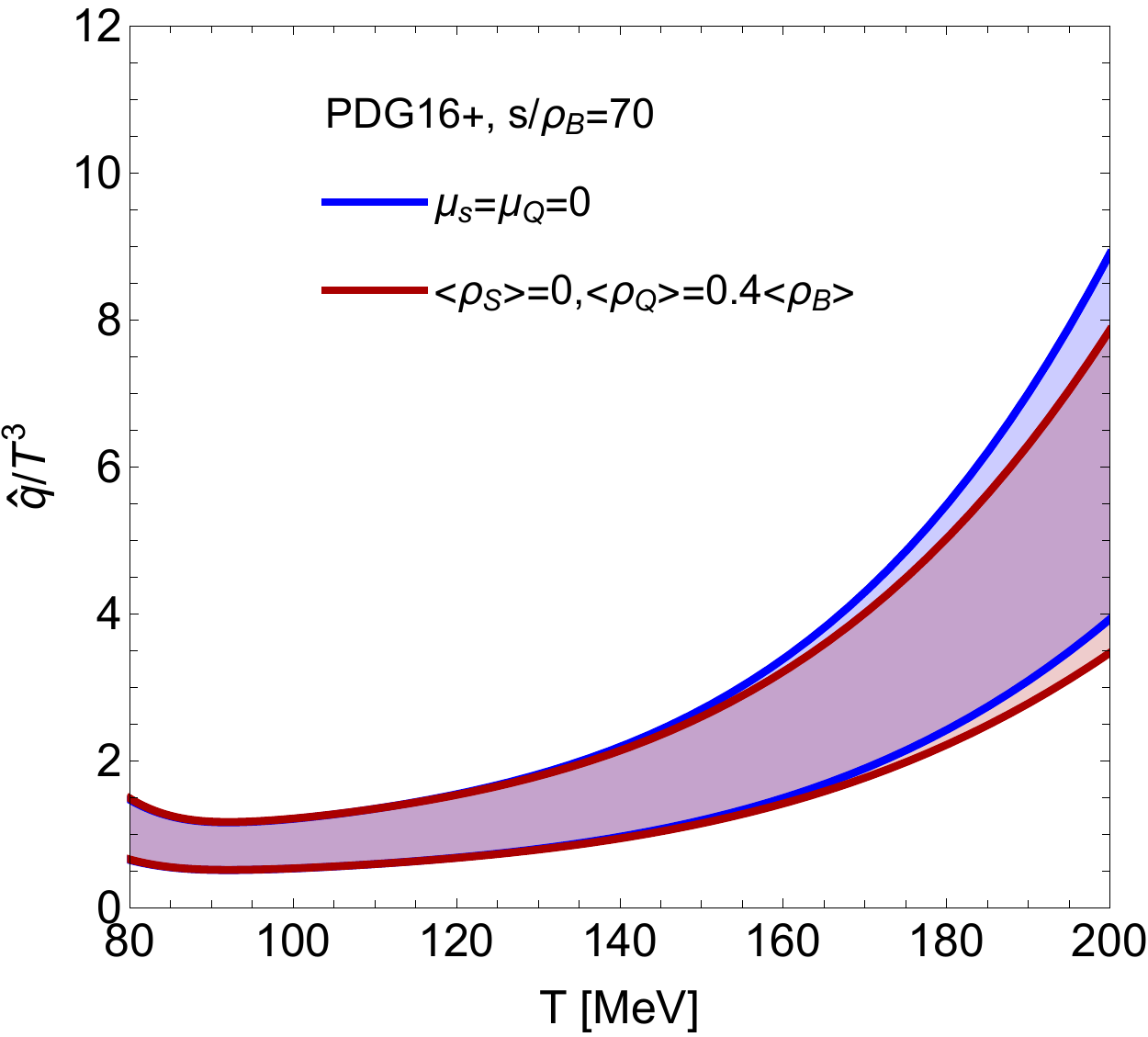}\\
    \includegraphics[width=\linewidth]{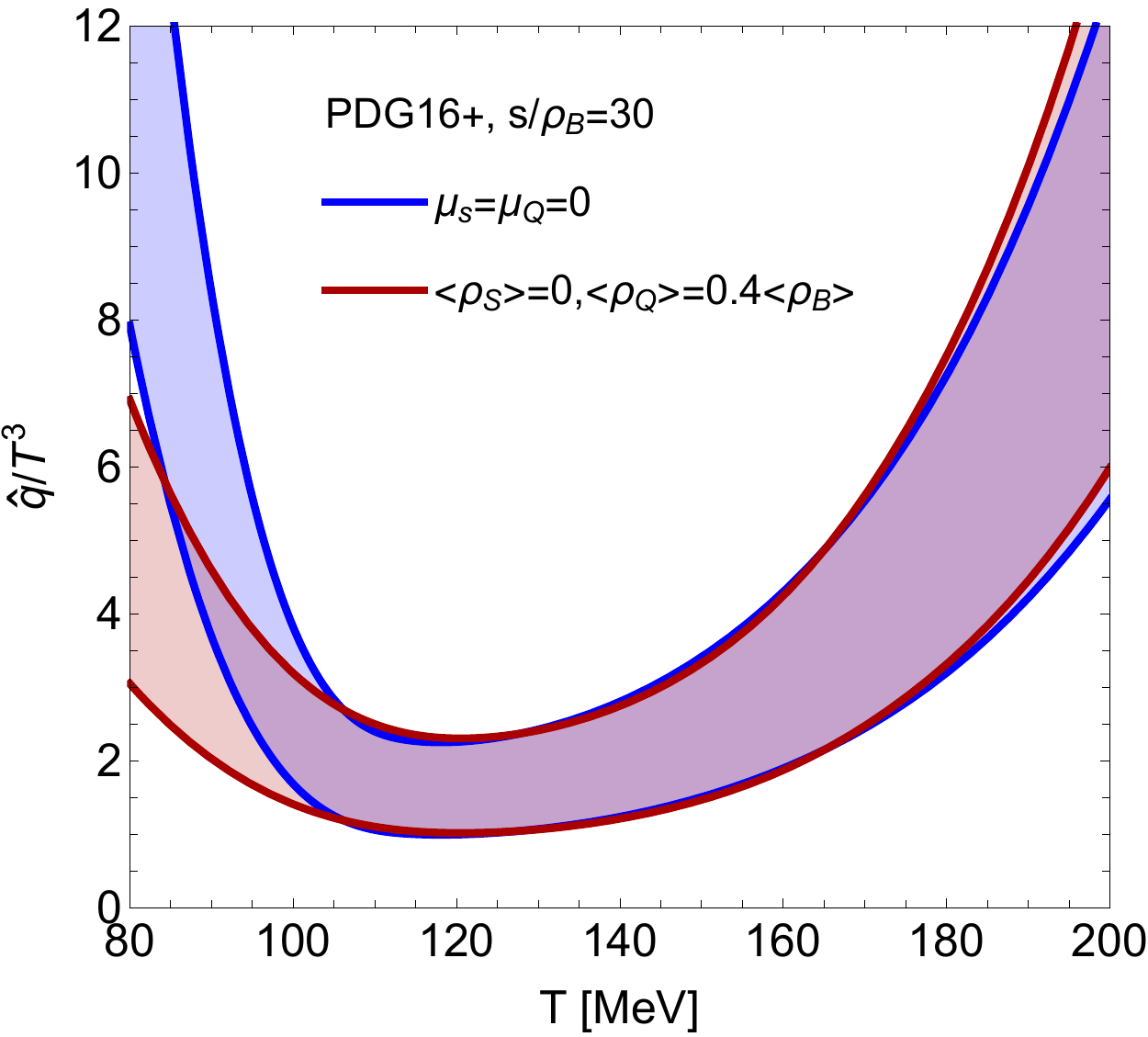}
    \caption{$\hat{q}(T,\mu_B)/T^3$  along isentropes assuming either a 2D EoS where $\mu_S=\mu_Q=0$ (blue) or a 4D EoS where strangeness neutrality and the ratio to protons/neutrons is simposed (red).}\label{fig:sn}
\end{figure}
First we consider a beam energy of $\sqrt{s_{NN}}=27$ GeV, where the entropy per baryon number at freeze-out is $s/\rho_B=70$, as shown in Fig.\ \ref{fig:sn} (a). At small enough $\mu_B$ there is no effect from the inclusion of strangeness neutrality on $\hat{q}/T^3$.  Thus, we do not consider higher beam energies.  However, once one reaches low enough beam energies such as $\sqrt{s_{NN}}<14.5$ GeV where  $s/\rho_B=30$, one can see that the impact from the 4D EoS plays some role. However, the biggest role is played by the isentrope itself, that gives a bend in $\hat{q}/T^3$ due to the assumption that entropy is conserved throughout the evolution of the system.

\section{Results: Testing $\eta T/w\sim T^3/\hat{q}$}

In Ref.~\cite{Majumder:2007zh}, a quasi-empirical formula was established, which relates the shear viscosity to the inverse of $\hat{q}$ such that
\begin{equation}
    \eta T/w\sim T^3/\hat{q}.
\end{equation}
In the absence of direct calculations of transport coefficients from QCD, we must turn to phenomenological approaches, as performed in this paper. 
Here we test our two approaches for calculating $\eta T/w$ and  $\hat{q}/T^3$ in the hadron resonance gas model against each other, to see if this relationship leads to reasonable values of $\eta T/w$.

\begin{figure}
    \centering
    \includegraphics[width=\linewidth]{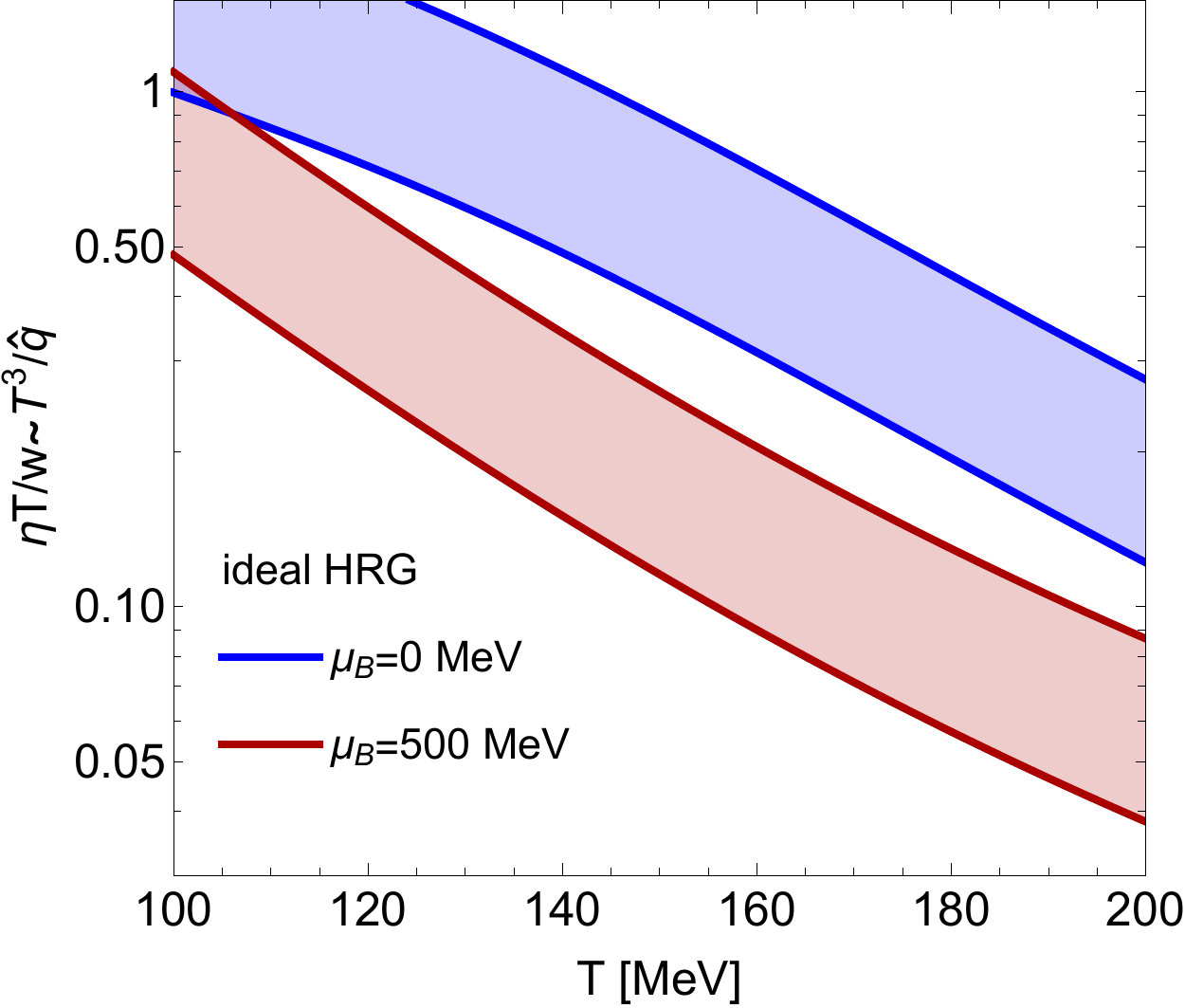}\\
    \includegraphics[width=\linewidth]{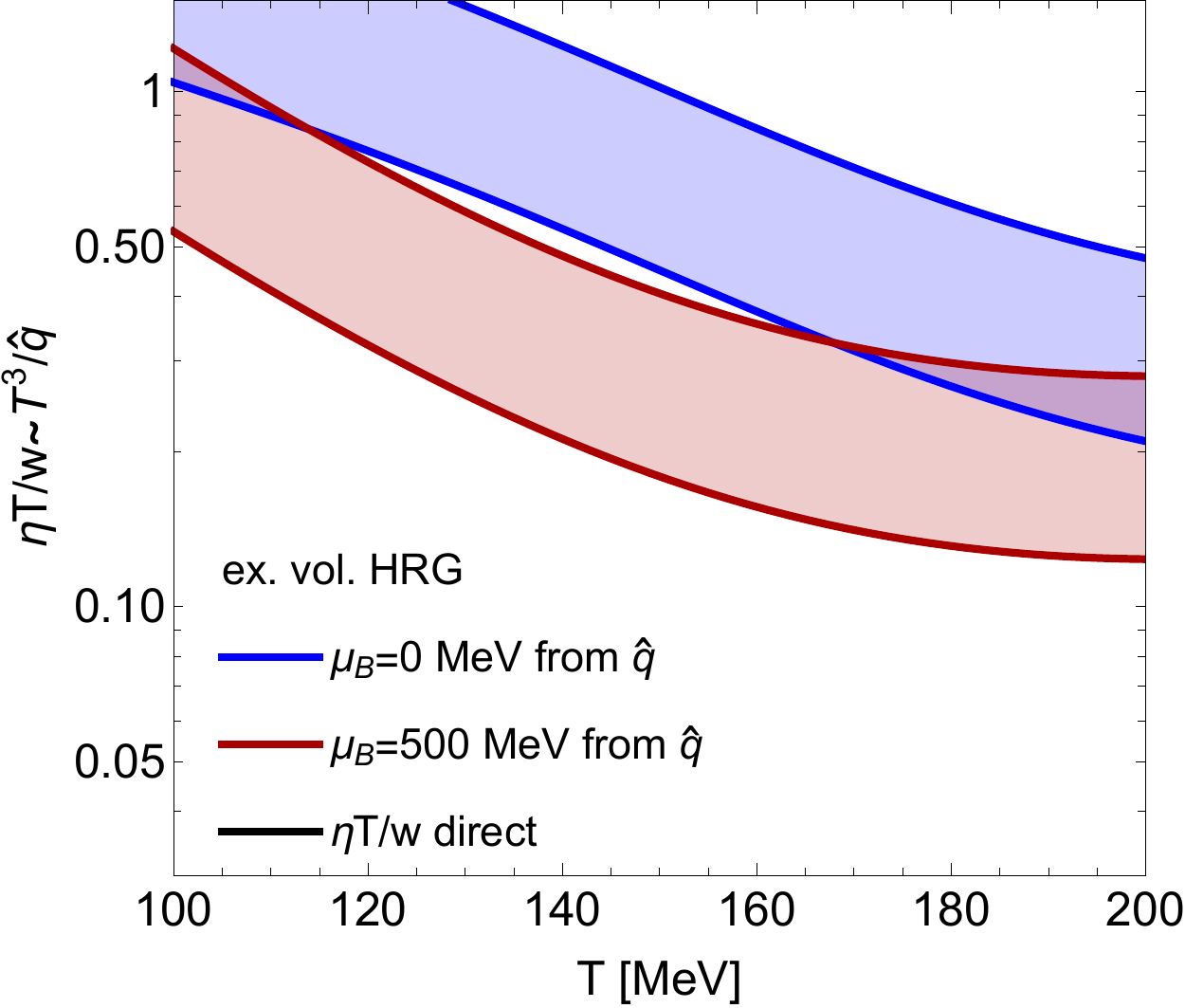}
    \caption{$\eta T/w \sim T^3/\hat{q}$  at $\mu_B=0$ (blue bands) and $\mu_B=500$ MeV (red bands) for an ideal HRG (top) and an excluded volume approach with $r=0.25$ fm from Ref.~\cite{McLaughlin:2021dph} (bottom). The solid black curve is $\eta T/w$ calculated directly in the excluded volume approach from Ref.~\cite{McLaughlin:2021dph} at $\mu_B=0$.  }\label{fig:etas} 
\end{figure}

The results of the extracted $\eta T/w$ from $\hat{q}$ are shown in Fig.\ \ref{fig:etas} and their magnitudes appear to be reasonable at $\mu_B=0$.  They have a value of $\eta T/w\sim 0.2$ for an ideal hadron resonance gas and around $\eta T/w\sim 0.3$ for the excluded volume approach (assuming $T_{pc}=200$ MeV). Additionally, at low temperatures $\eta T/w$ increases, which is to be expected.  An issue arises when one compares an ideal hadron resonance gas to the excluded volume one. If one uses an excluded volume approach, as was done in Refs.~\cite{Rischke:1991ke,Gorenstein:2007mw,NoronhaHostler:2012ug}, one finds that interactions reduce $\eta T/w$, which is in contrast to what we find using the $\hat{q}/T^3$ formalism from Ref.~\cite{Burke:2013yra}. Additionally, we can  compare the direct result from an excluded volume calculation from Ref.~\cite{McLaughlin:2021dph} (assuming the same radius in both calculations of $r=0.25$ fm) and we see that, while they somewhat match at large temperatures, they deviate at low temperatures.  

\begin{figure}
    \centering
    \includegraphics[width=\linewidth]{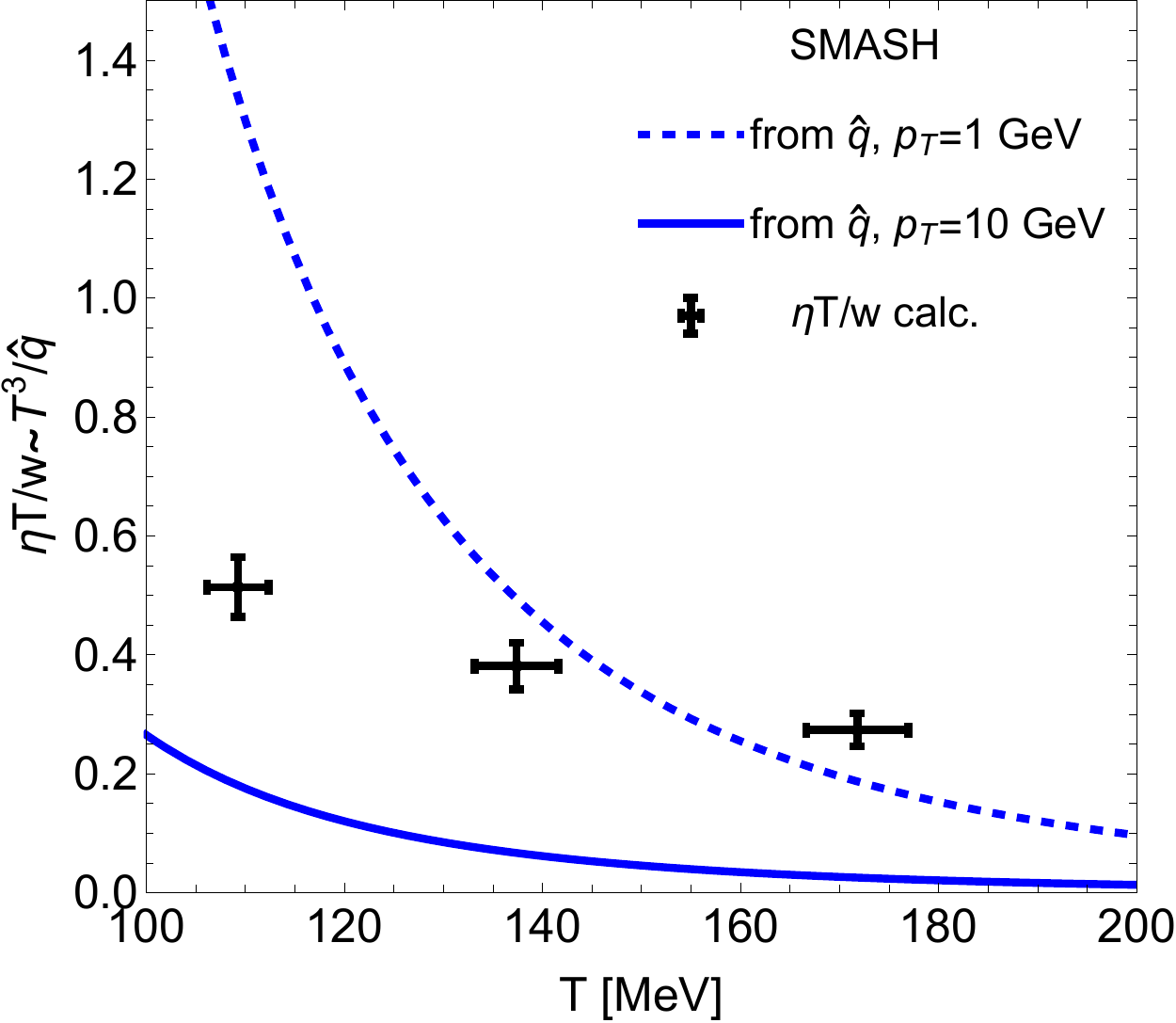}
    \caption{Comparison of the direct $\eta T/w$ calculation from Ref.~\cite{Rose:2017bjz}  to the relationship $\eta T/w \sim T^3/\hat{q}$ \cite{Dorau:2019ozd} at $\mu_B=0$ from SMASH}\label{fig:SMASH}
\end{figure}
Because of the issue with the relationship between $\eta T/w$ and $\hat{q}$, we looked at recent calculations from the hadronic transport code SMASH for guidance. The code has been used to calculate both $\eta T/w$  \cite{Rose:2017bjz} and $\tilde{q}$  \cite{Dorau:2019ozd}, that we then convert into $\hat{q}/T^3$. We compare the direct $\eta T/w$ calculation to $\eta T/w \sim T^3/\hat{q}$ in Fig.\ \ref{fig:SMASH}. We find a somewhat similar result as our HRG model calculation if we consider $\tilde{q}$ for $p_T=1$ GeV. The relationship $\eta T/w \sim T^3/\hat{q}$ approximately holds at high temperatures, while deviations arise at low temperatures. However, if one considers a larger $p_T$, the estimated $\eta T/w $ is significantly smaller than the directly calculated one. Since both the HRG model and hadronic transport cannot consistently reproduce $\eta T/w \sim T^3/\hat{q}$, further investigations into this mismatch within a hadron resonance gas model might be needed.

Finally, we also explore $\eta T/w \sim T^3/\hat{q}$ at finite $\mu_B$. Qualitatively, we find that the extracted $\eta T/w $ at $\mu_B=500$ MeV behaves as expected, 
as it is smaller at large $\mu_B$ at a fixed temperature. We observe that the excluded volume approach demonstrates less sensitivity to $\mu_B$ compared to the ideal HRG model case.

\section{Conclusions}

In this paper we calculate $\hat{q}/T^3$ within the hadron resonance gas model formalism for both an ideal gas of hadrons and an excluded volume approach. Additionally, if we incorporate the experimental uncertainty into that calculation and assume a higher pseudo critical temperature for the maximum of $\hat{q}/T^3$, we are able to smoothly connect our results to those of the JET collaboration. At the time of finalizing this paper JETSCAPE released new $\hat{q}/T^3$ results that would also smoothly match to our hadronic results \cite{Cao:2021keo}.  Already a significant effort has been invested to determine this coefficient within the Quark Gluon Plasma phase \cite{Baier:1996sk,CaronHuot:2008ni,Muller:2012hr,Zhang:2012jd,Panero:2013pla,Xu:2014tda,Majumder:2014gda,Majumder:2012sh,Mueller:2016gko,Andres:2016iys,Rapp:2018qla,Kumar:2019uvu,Kumar:2020wvb}.  Our results indicate that dynamical models should also study the effect of energy loss more thoroughly within the hadron gas phase.

Furthermore, we find that within hadronic models (both for an excluded volume hadron resonance gas and hadronic transport- SMASH) it is not clear whether the relationship $\eta T/w \sim T^3/\hat{q}$  holds, which warrants further investigation. 

Finally, we find an enhancement of $\hat{q}/T^3$ at finite $\mu_B$.  At large $\mu_B$ we find that the results in an excluded volume approach deviate more strongly from the ideal hadron resonance gas ones (which is expected). While we anticipate a low chance of jets being produced at the beam energy scan (although we do not exclude the possibility of mini-jets/low $p_T$ jets), our results demonstrate that any jets produced would be heavily suppressed.

\section{Acknowledgements }

 J.N.H. acknowledges the support from the US-DOE Nuclear Science Grant No. DE-SC0020633. E.M. was been supported by the
National Science Foundation via grant PHY-1560077. The authors also acknowledge support from the Illinois Campus Cluster, a computing resource that is operated by the Illinois Campus Cluster Program (ICCP) in conjunction with the National Center for Supercomputing Applications (NCSA), and which is supported by funds from the University of Illinois at Urbana-Champaign.
This material is based upon work supported by the National Science Foundation under grant
no. PHY-1654219 and by the U.S. Department of Energy, Office of Science,
Office of Nuclear Physics, within the framework of the Beam Energy Scan Theory (BEST) Topical
Collaboration. P.P.  acknowledges support by the DFG grant SFB/TR55.

\nocite{*}
\bibliography{all}

\begin{thebibliography}{10}

\bibitem{Adil:2005qn}
A.~Adil and M.~Gyulassy, ``{3D jet tomography of twisted strongly coupled quark
  gluon plasmas},'' {\em Phys. Rev. C}, vol.~72, p.~034907, 2005.

\bibitem{Andres:2019eus}
C.~Andres, N.~Armesto, H.~Niemi, R.~Paatelainen, and C.~A. Salgado, ``{Jet
  quenching as a probe of the initial stages in heavy-ion collisions},'' {\em
  Phys. Lett. B}, vol.~803, p.~135318, 2020.

\bibitem{Connors:2017ptx}
M.~Connors, C.~Nattrass, R.~Reed, and S.~Salur, ``{Jet measurements in heavy
  ion physics},'' {\em Rev. Mod. Phys.}, vol.~90, p.~025005, 2018.

\bibitem{Burke:2013yra}
K.~M. Burke {\em et~al.}, ``{Extracting the jet transport coefficient from jet
  quenching in high-energy heavy-ion collisions},'' {\em Phys. Rev. C},
  vol.~90, no.~1, p.~014909, 2014.

\bibitem{Aoki:2006br}
Y.~Aoki, Z.~Fodor, S.~Katz, and K.~Szabo, ``{The QCD transition temperature:
  Results with physical masses in the continuum limit},'' {\em Phys. Lett. B},
  vol.~643, pp.~46--54, 2006.

\bibitem{Borsanyi:2010cj}
S.~Borsanyi, G.~Endrodi, Z.~Fodor, A.~Jakovac, S.~D. Katz, S.~Krieg, C.~Ratti,
  and K.~K. Szabo, ``{The QCD equation of state with dynamical quarks},'' {\em
  JHEP}, vol.~11, p.~077, 2010.

\bibitem{Borsanyi:2010bp}
S.~Borsanyi, Z.~Fodor, C.~Hoelbling, S.~D. Katz, S.~Krieg, C.~Ratti, and K.~K.
  Szabo, ``{Is there still any $T_c$ mystery in lattice QCD? Results with
  physical masses in the continuum limit III},'' {\em JHEP}, vol.~09, p.~073,
  2010.

\bibitem{Bazavov:2014pvz}
A.~Bazavov {\em et~al.}, ``{Equation of state in ( 2+1 )-flavor QCD},'' {\em
  Phys. Rev. D}, vol.~90, p.~094503, 2014.

\bibitem{Bazavov:2017tot}
A.~Bazavov {\em et~al.}, ``{Skewness and kurtosis of net baryon-number
  distributions at small values of the baryon chemical potential},'' {\em Phys.
  Rev. D}, vol.~96, no.~7, p.~074510, 2017.

\bibitem{Bazavov:2017dus}
A.~Bazavov {\em et~al.}, ``{The QCD Equation of State to $\mathcal{O}(\mu_B^6)$
  from Lattice QCD},'' {\em Phys. Rev. D}, vol.~95, no.~5, p.~054504, 2017.

\bibitem{Mukherjee:2019eou}
S.~Mukherjee and V.~Skokov, ``{Universality driven analytic structure of QCD
  crossover: radius of convergence in baryon chemical potential},'' 9 2019.

\bibitem{Borsanyi:2020fev}
S.~Borsanyi, Z.~Fodor, J.~N. Guenther, R.~Kara, S.~D. Katz, P.~Parotto,
  A.~Pasztor, C.~Ratti, and K.~K. Szabo, ``{QCD Crossover at Finite Chemical
  Potential from Lattice Simulations},'' {\em Phys. Rev. Lett.}, vol.~125,
  no.~5, p.~052001, 2020.

\bibitem{Bellwied:2013cta}
R.~Bellwied, S.~Borsanyi, Z.~Fodor, S.~D. Katz, and C.~Ratti, ``{Is there a
  flavor hierarchy in the deconfinement transition of QCD?},'' {\em Phys. Rev.
  Lett.}, vol.~111, p.~202302, 2013.

\bibitem{Bellwied:2019pxh}
R.~Bellwied, S.~Borsanyi, Z.~Fodor, J.~N. Guenther, J.~Noronha-Hostler,
  P.~Parotto, A.~Pasztor, C.~Ratti, and J.~M. Stafford, ``{Off-diagonal
  correlators of conserved charges from lattice QCD and how to relate them to
  experiment},'' {\em Phys. Rev. D}, vol.~101, no.~3, p.~034506, 2020.

\bibitem{Noronha-Hostler:2016rpd}
J.~Noronha-Hostler, R.~Bellwied, J.~Gunther, P.~Parotto, A.~Pasztor,
  I.~Portillo~Vazquez, and C.~Ratti, ``{Kaon fluctuations from lattice QCD},''
  7 2016.

\bibitem{Bellwied:2018tkc}
R.~Bellwied, J.~Noronha-Hostler, P.~Parotto, I.~Portillo~Vazquez, C.~Ratti, and
  J.~M. Stafford, ``{Freeze-out temperature from net-kaon fluctuations at
  energies available at the BNL Relativistic Heavy Ion Collider},'' {\em Phys.
  Rev. C}, vol.~99, no.~3, p.~034912, 2019.

\bibitem{Bluhm:2018aei}
M.~Bluhm and M.~Nahrgang, ``{Freeze-out conditions from strangeness observables
  at RHIC},'' {\em Eur. Phys. J. C}, vol.~79, no.~2, p.~155, 2019.

\bibitem{Alba:2020jir}
P.~Alba, V.~M. Sarti, J.~Noronha-Hostler, P.~Parotto, I.~Portillo-Vazquez,
  C.~Ratti, and J.~Stafford, ``{Influence of hadronic resonances on the
  chemical freeze-out in heavy-ion collisions},'' {\em Phys. Rev. C}, vol.~101,
  no.~5, p.~054905, 2020.

\bibitem{Flor:2020fdw}
F.~A. Flor, G.~Olinger, and R.~Bellwied, ``{Flavor and Energy Dependence of
  Chemical Freeze-out Temperatures in Relativistic Heavy Ion Collisions from
  RHIC-BES to LHC Energies},'' 9 2020.

\bibitem{Rougemont:2017tlu}
R.~Rougemont, R.~Critelli, J.~Noronha-Hostler, J.~Noronha, and C.~Ratti,
  ``{Dynamical versus equilibrium properties of the QCD phase transition: A
  holographic perspective},'' {\em Phys. Rev. D}, vol.~96, no.~1, p.~014032,
  2017.

\bibitem{Critelli:2017oub}
R.~Critelli, J.~Noronha, J.~Noronha-Hostler, I.~Portillo, C.~Ratti, and
  R.~Rougemont, ``{Critical point in the phase diagram of primordial
  quark-gluon matter from black hole physics},'' {\em Phys. Rev. D}, vol.~96,
  no.~9, p.~096026, 2017.

\bibitem{1848212}
J.~Grefa, J.~Noronha, J.~Noronha-Hostler, I.~Portillo, C.~Ratti, and
  R.~Rougemont, ``{Hot and dense quark-gluon plasma thermodynamics from
  holographic black holes},'' 2 2021.

\bibitem{Majumder:2007zh}
A.~Majumder, B.~Muller, and X.-N. Wang, ``{Small shear viscosity of a
  quark-gluon plasma implies strong jet quenching},'' {\em Phys. Rev. Lett.},
  vol.~99, p.~192301, 2007.

\bibitem{Wang:2009qb}
W.-t. Deng and X.-N. Wang, ``{Multiple Parton Scattering in Nuclei: Modified
  DGLAP Evolution for Fragmentation Functions},'' {\em Phys. Rev. C}, vol.~81,
  p.~024902, 2010.

\bibitem{Airapetian:2007vu}
A.~Airapetian {\em et~al.}, ``{Hadronization in semi-inclusive deep-inelastic
  scattering on nuclei},'' {\em Nucl. Phys. B}, vol.~780, pp.~1--27, 2007.

\bibitem{Aaron:2009aa}
F.~Aaron {\em et~al.}, ``{Combined Measurement and QCD Analysis of the
  Inclusive e+- p Scattering Cross Sections at HERA},'' {\em JHEP}, vol.~01,
  p.~109, 2010.

\bibitem{Alba:2017mqu}
P.~Alba {\em et~al.}, ``{Constraining the hadronic spectrum through QCD
  thermodynamics on the lattice},'' {\em Phys. Rev. D}, vol.~96, no.~3,
  p.~034517, 2017.

\bibitem{Betz:2016ayq}
B.~Betz, M.~Gyulassy, M.~Luzum, J.~Noronha, J.~Noronha-Hostler, I.~Portillo,
  and C.~Ratti, ``{Cumulants and nonlinear response of high $p_T$ harmonic flow
  at $\sqrt{s_{NN}}=5.02$ TeV},'' {\em Phys. Rev. C}, vol.~95, no.~4,
  p.~044901, 2017.

\bibitem{Nahrgang:2016lst}
M.~Nahrgang, J.~Aichelin, P.~B. Gossiaux, and K.~Werner, ``{Toward a consistent
  evolution of the quark-gluon plasma and heavy quarks},'' {\em Phys. Rev. C},
  vol.~93, no.~4, p.~044909, 2016.

\bibitem{Prado:2016szr}
C.~A. Prado, J.~Noronha-Hostler, R.~Katz, A.~A. Suaide, J.~Noronha, M.~G.
  Munhoz, and M.~R. Cosentino, ``{Event-by-event correlations between soft
  hadrons and $D^0$ mesons in 5.02 TeV PbPb collisions at the CERN Large Hadron
  Collider},'' {\em Phys. Rev. C}, vol.~96, no.~6, p.~064903, 2017.

\bibitem{Katz:2019fkc}
R.~Katz, C.~A. Prado, J.~Noronha-Hostler, J.~Noronha, and A.~A. Suaide,
  ``{Sensitivity study with a D and B mesons modular simulation code of heavy
  flavor RAA and azimuthal anisotropies based on beam energy, initial
  conditions, hadronization, and suppression mechanisms},'' {\em Phys. Rev. C},
  vol.~102, no.~2, p.~024906, 2020.

\bibitem{Dorau:2019ozd}
P.~Dorau, J.-B. Rose, D.~Pablos, and H.~Elfner, ``{Jet Quenching in the Hadron
  Gas: An Exploratory Study},'' {\em Phys. Rev. C}, vol.~101, no.~3, p.~035208,
  2020.

\bibitem{McLaughlin:2021dph}
E.~McLaughlin, J.~Rose, T.~Dore, P.~Parotto, C.~Ratti, and J.~Noronha-Hostler,
  ``{Building a testable shear viscosity across the QCD phase diagram},'' 3
  2021.

\bibitem{Auvinen:2013sba}
J.~Auvinen and H.~Petersen, ``{Evolution of elliptic and triangular flow as a
  function of $\sqrt{{s}_{NN}}$ in a hybrid model},'' {\em Phys. Rev. C},
  vol.~88, no.~6, p.~064908, 2013.

\bibitem{Adamczyk:2017nof}
L.~Adamczyk {\em et~al.}, ``{Beam Energy Dependence of Jet-Quenching Effects in
  Au+Au Collisions at $\sqrt{s_{_{ \mathrm{NN}}}}$ = 7.7, 11.5, 14.5, 19.6, 27,
  39, and 62.4 GeV},'' {\em Phys. Rev. Lett.}, vol.~121, no.~3, p.~032301,
  2018.

\bibitem{Noronha-Hostler:2019ayj}
J.~Noronha-Hostler, P.~Parotto, C.~Ratti, and J.~Stafford, ``{Lattice-based
  equation of state at finite baryon number, electric charge and strangeness
  chemical potentials},'' {\em Phys. Rev. C}, vol.~100, no.~6, p.~064910, 2019.

\bibitem{Shen:2017ruz}
C.~Shen, G.~Denicol, C.~Gale, S.~Jeon, A.~Monnai, and B.~Schenke, ``{A hybrid
  approach to relativistic heavy-ion collisions at the RHIC BES energies},''
  {\em Nucl. Phys. A}, vol.~967, pp.~796--799, 2017.

\bibitem{Feng:2018anl}
B.~Feng, C.~Greiner, S.~Shi, and Z.~Xu, ``{Viscous effects on the dynamical
  evolution of QCD matter during the first-order confinement phase transition
  in heavy-ion collisions},'' {\em Phys. Lett. B}, vol.~782, pp.~262--267,
  2018.

\bibitem{Fotakis:2019nbq}
J.~A. Fotakis, M.~Greif, C.~Greiner, G.~S. Denicol, and H.~Niemi, ``{Diffusion
  processes involving multiple conserved charges: A study from kinetic theory
  and implications to the fluid-dynamical modeling of heavy ion collisions},''
  {\em Phys. Rev. D}, vol.~101, no.~7, p.~076007, 2020.

\bibitem{Martinez:2019rlp}
M.~Martinez, M.~D. Sievert, D.~E. Wertepny, and J.~Noronha-Hostler, ``{Toward
  Initial Conditions of Conserved Charges Part II: The ICCING Monte Carlo
  Algorithm},'' 11 2019.

\bibitem{Martinez:2019jbu}
M.~Martinez, M.~D. Sievert, D.~E. Wertepny, and J.~Noronha-Hostler, ``{Initial
  state fluctuations of QCD conserved charges in heavy-ion collisions},'' 11
  2019.

\bibitem{Dore:2020jye}
T.~Dore, J.~Noronha-Hostler, and E.~McLaughlin, ``{Far-from-equilibrium search
  for the QCD critical point},'' {\em Phys. Rev. D}, vol.~102, no.~7,
  p.~074017, 2020.

\bibitem{Rischke:1991ke}
D.~H. Rischke, M.~I. Gorenstein, H.~Stoecker, and W.~Greiner, ``{Excluded
  volume effect for the nuclear matter equation of state},'' {\em Z. Phys. C},
  vol.~51, pp.~485--490, 1991.

\bibitem{Gorenstein:2007mw}
M.~Gorenstein, M.~Hauer, and O.~Moroz, ``{Viscosity in the excluded volume
  hadron gas model},'' pp.~214--220, 8 2007.

\bibitem{NoronhaHostler:2012ug}
J.~Noronha-Hostler, J.~Noronha, and C.~Greiner, ``{Hadron Mass Spectrum and the
  Shear Viscosity to Entropy Density Ratio of Hot Hadronic Matter},'' {\em
  Phys. Rev. C}, vol.~86, p.~024913, 2012.

\bibitem{Rose:2017bjz}
J.~B. Rose, J.~Torres-Rincon, A.~Sch\"afer, D.~Oliinychenko, and H.~Petersen,
  ``{Shear viscosity of a hadron gas and influence of resonance lifetimes on
  relaxation time},'' {\em Phys. Rev. C}, vol.~97, no.~5, p.~055204, 2018.

\bibitem{Cao:2021keo}
S.~Cao {\em et~al.}, ``{Determining the jet transport coefficient $\hat{q}$
  from inclusive hadron suppression measurements using Bayesian parameter
  estimation},'' 2 2021.

\bibitem{Baier:1996sk}
R.~Baier, Y.~L. Dokshitzer, A.~H. Mueller, S.~Peigne, and D.~Schiff,
  ``{Radiative energy loss and p(T) broadening of high-energy partons in
  nuclei},'' {\em Nucl. Phys. B}, vol.~484, pp.~265--282, 1997.

\bibitem{CaronHuot:2008ni}
S.~Caron-Huot, ``{O(g) plasma effects in jet quenching},'' {\em Phys. Rev. D},
  vol.~79, p.~065039, 2009.

\bibitem{Muller:2012hr}
B.~M\"uller, ``{No Pain, No Gain: Hard Probes of the Quark-Gluon Plasma Coming
  of Age},'' {\em Nucl. Phys. A}, vol.~910-911, pp.~5--11, 2013.

\bibitem{Zhang:2012jd}
Z.-q. Zhang, D.-f. Hou, and H.-c. Ren, ``{The finite 't Hooft coupling
  correction on jet quenching parameter in a $\mathcal N=4$ Super Yang-Mills
  Plasma},'' {\em JHEP}, vol.~01, p.~032, 2013.

\bibitem{Panero:2013pla}
M.~Panero, K.~Rummukainen, and A.~Sch\"afer, ``{Lattice Study of the Jet
  Quenching Parameter},'' {\em Phys. Rev. Lett.}, vol.~112, no.~16, p.~162001,
  2014.

\bibitem{Xu:2014tda}
J.~Xu, J.~Liao, and M.~Gyulassy, ``{Consistency of Perfect Fluidity and Jet
  Quenching in semi-Quark-Gluon Monopole Plasmas},'' {\em Chin. Phys. Lett.},
  vol.~32, no.~9, p.~092501, 2015.

\bibitem{Majumder:2014gda}
A.~Majumder and J.~Putschke, ``{Mass depletion: a new parameter for
  quantitative jet modification},'' {\em Phys. Rev. C}, vol.~93, no.~5,
  p.~054909, 2016.

\bibitem{Majumder:2012sh}
A.~Majumder, ``{Calculating the jet quenching parameter $\hat{q}$ in lattice
  gauge theory},'' {\em Phys. Rev. C}, vol.~87, p.~034905, 2013.

\bibitem{Mueller:2016gko}
A.~Mueller, B.~Wu, B.-W. Xiao, and F.~Yuan, ``{Probing Transverse Momentum
  Broadening in Heavy Ion Collisions},'' {\em Phys. Lett. B}, vol.~763,
  pp.~208--212, 2016.

\bibitem{Andres:2016iys}
C.~Andr\'es, N.~Armesto, M.~Luzum, C.~A. Salgado, and P.~Zurita, ``{Energy
  versus centrality dependence of the jet quenching parameter $\hat{q}$ at RHIC
  and LHC: a new puzzle?},'' {\em Eur. Phys. J. C}, vol.~76, no.~9, p.~475,
  2016.

\bibitem{Rapp:2018qla}
A.~Beraudo {\em et~al.}, ``{Extraction of Heavy-Flavor Transport Coefficients
  in QCD Matter},'' {\em Nucl. Phys. A}, vol.~979, pp.~21--86, 2018.

\bibitem{Kumar:2019uvu}
A.~Kumar, A.~Majumder, and C.~Shen, ``{Energy and scale dependence of $\hat{q}$
  and the \textquotedblleft{}JET puzzle\textquotedblright{}},'' {\em Phys. Rev.
  C}, vol.~101, no.~3, p.~034908, 2020.

\bibitem{Kumar:2020wvb}
A.~Kumar, A.~Majumder, and J.~H. Weber, ``{Jet transport coefficient $\hat{q}$
  in (2+1)-flavor lattice QCD},'' 10 2020.

\end{thebibliography}
\bibliographystyle{ieeetr}

\end{document}